\documentclass{IEEEtran}

\usepackage{hyperref}

\usepackage{cite}
\usepackage{amsmath,amssymb,amsfonts}
\usepackage{graphicx}
\usepackage{xurl}
\usepackage{textcomp,nicefrac}

\usepackage{listings}
\usepackage{xcolor}
\usepackage{tabularx}

\definecolor{codegreen}{rgb}{0,0.4,0}
\definecolor{codegray}{rgb}{0.5,0.5,0.5}
\definecolor{codepurple}{rgb}{0.48,0,0.70}
\definecolor{backcolour}{rgb}{0.96,0.96,0.93}

\lstdefinestyle{mystyle}{
    basewidth=4.3pt,
    backgroundcolor=\color{backcolour},   
    commentstyle=\color{codegreen},
    keywordstyle=\color{magenta},
    numberstyle=\tiny\color{codegray},
    stringstyle=\color{codepurple},
    basicstyle=\ttfamily\footnotesize,
    breakatwhitespace=false,         
    breaklines=true,                 
    captionpos=b,                    
    keepspaces=true,                 
    numbers=left,                    
    numbersep=5pt,                  
    showspaces=false,                
    showstringspaces=false,
    showtabs=false,                  
    tabsize=2
}

\def\BibTeX{{\rm B\kern-.05em{\sc i\kern-.025em b}\kern-.08em
T\kern-.1667em\lower.7ex\hbox{E}\kern-.125emX}}

\begin{document}

\title{OpenIPMC: a free and open source Intelligent Platform Management Controller Software}

\author{{\large Luigi Calligaris, André Cascadan, Luis E. Ardila-Perez, Bruno Casu, Alison França da Costa, Ailton Akira Shinoda, Lucas Arruda Ramalho and Oliver Sander}

\thanks{
L. Calligaris (corresponding author, email: luigi.calligaris at cern.ch), A. Cascadan and B. Casu are with the Scientific Computing Center (NCC) of São Paulo State University (UNESP), Rua Dr. Bento Teobaldo Ferraz, 271, São Paulo - SP, 01140-070, Brazil.

A. França da Costa and A. A. Shinoda are with the Electrical Engineering Department (FEIS) of São Paulo State University (UNESP), Av. Professor José Carlos Rossi, 1370 Campus III, Ilha Solteira - SP, 15385-000, Brazil.

L. A. Ramalho is with Exact and Earth Sciences Department (FACET) of Mato Grosso State University (UNEMAT), Rua A, S/n, Bairro São Raimundo, 78390-000, Caixa Postal 92, Barra do Bugres, Mato Grosso.

L. E. Ardila-Perez and O. Sander are with the Institute for Data Processing and Electronics (IPE) of Karlsruhe Institute of Technology, Hermann-von-Helmholtz-Platz 1, D-76344 Eggenstein-Leopoldshafen, Germany.

This work is supported by the Fundação de Amparo à Pesquisa do Estado de São Paulo (FAPESP), through grants number 18/18955-0 and 17/16245-3.
}
}

\maketitle

\begin{abstract}
OpenIPMC is a free and open source software designed to implement the logic of an Intelligent Platform Management Controller (IPMC). An IPMC is a fundamental component of electronic boards conformant to the Advanced Telecommunications Computing Architecture (ATCA) standard, currently being adopted by a number of high energy physics experiments. The IPMC is responsible for monitoring the health parameters of the board, managing its power states, and providing board control, debug and recovery functions to remote clients. OpenIPMC is based on the FreeRTOS real-time operating system and is designed to be architecture-independent, allowing it to be used in firmware designed for a variety of microcontrollers. Having a fully free and open source code is an innovative aspect for this kind of software, enabling full customization by the user. In this work we present the features and structure of OpenIPMC as well as its example implementations on Xilinx Zynq UltraScale+ (ZynqUS+), Espressif ESP32 and ST Microelectronics STM32 architectures.
\end{abstract}

\begin{IEEEkeywords}
PICMG, ATCA, IPMC, Electronic board management
\end{IEEEkeywords}

\vspace{-2mm}
\section{Introduction}
\label{sec:intro}

The Advanced Telecommunications Computing Architecture (ATCA) standard \cite{picmg_3_0} is developed by a consortium of leading computer hardware manufacturers known as the PCI Industrial Computer Manufacturing Group (PICMG) \cite{picmg_site}. This standard defines mechanical, electrical and functional design rules, connector pin assignments and communication protocols to be used in the design of electronic boards and their housing shelves for industrial computing applications. The rules aim to guarantee a high availability and reliability of the deployed systems, an objective which is achieved with the aid of a sophisticated Hardware Platform Management (HPM) \cite{HPM} infrastructure. The ATCA standard is widely adopted in the telecommunications industry and its use extends to a broader range of applications, such as medical equipment\cite{atca_pet} secure networking, military electronics and large physics experiments \cite{picmg_site_atca}.

The focus on high availability and reliability, the large data bandwidth offered by the shelf backplane, the availability of large electrical power, good thermal dissipation, and the possibility to insert and remove boards and other components in a running system (\emph{hot swap}) make ATCA systems very attractive for use in high energy physics experiments \cite{CERN-LHCC-2017-009, CERN-LHCC-2017-014, CERN-LHCC-2017-021, CERN-LHCC-2017-005, CERN-LHCC-2017-020, CERN-LHCC-2014-001, CERN-LHCC-2014-016}, where the requirements for very large detector read-out rate, low latency, high availability and compact physical size of the back-end systems of the detectors push the limits of current technology. Examples of ATCA boards used in High Energy Physics are the Serenity \cite{Rose:2019oiy} and Apollo \cite{Albert:2019kvt} boards, which are going to be used in the back-end of the tracker detector of the CMS experiment \cite{Calligaris:2020unu} following its Phase-2 upgrade.

Each electronic board compliant to the ATCA standard is required to host an Intelligent Platform Management Controller (IPMC), which is typically implemented using a microcontroller running a firmware that implements the IPMC functions. Many IPMC solutions have been proposed over the years, some of which commercial in nature \cite{pigeon-point, cern-ipmc} and others non-commercial in nature \cite{lapp-ipmc, flash-ipmc, ramalho_ipmc_pulsa2b, paiva_thesis}. These solutions employ a firmware specifically written for their intended target microcontroller, consequently making it tedious to migrate the firmware in the event those parts become obsolete. Furthermore, vendor tool-chains supporting those older parts tend to be excluded from new software updates, making them rely on the support of legacy operating systems, which can be difficult to operate over the lifetime of the target boards. Lastly, closed-source implementations - including commercial ones - often pose significant bureaucratic barriers to developers by requiring their institutions to sign Non-Disclosure Agreements (NDA). This may cause lengthy approval processes by the home institution of the researchers and prevent non-staff participants (e.g. students) from participating officially to the project. These are some of the reasons that motivated us to look into developing a free and open-source solution.

\section{The ATCA Shelf}
An ATCA shelf is a standardized form factor chassis accommodating Field Replaceable Units (FRU). These can be various ``intelligent"  (i.e. capable of mutual coordination) components such as cooling fan trays, power supplies or user-designed electronic boards, which are the focus of our development. FRUs need to be compliant to a set of mechanical, electrical and interface specifications, as defined in the PICMG standard, to ensure their proper inter-operation. From the point of view of an ATCA electronic board, the resources made available by a shelf are:

\textbf{Power:} Each board is powered by a two-channel, redundant, -48 V rail.

\textbf{Cooling:} Redundant fan trays drive an air stream to remove the heat generated by the electronics.

\textbf{Data and clock bus:} ATCA specifies a number of backplane topologies, supporting the transmission of synchronization clock signals and high speed links between the boards.

\textbf{IPMB:} The backplane exposes to all FRUs a dual-redundant two-wire bus, compatible with the $\text{I}^2\text{C}$ protocol with a signalling level of 3.3V. The two buses are named Intelligent Platform Management Bus A and B (IPMB-A and IPMB-B), also referenced collectively as IPMB-0. The two buses are used by the FRUs and the Shelf Management Controller (ShMC) to relay messages in multi-master mode, that is, listening as I$^{2}$C slaves for messages addressed to them and taking control of one of the available buses as I$^{2}$C masters to send a message, managing failures and collisions in case they take place.

\textbf{Other management signals:} A set of pins in the backplane electrically encodes the address of the physical slot a board is inserted in.

\subsection{Hardware Platform Management}
The main task of the HPM system is monitoring the health of the hardware by collecting sensor data (voltages, current draws, temperatures, fan speeds, etc.) and taking corrective actions (increasing the fan speed, switching off power, trigger alarms, etc.) in case the measurements lie outside the nominal range. The system is also tasked with orchestrating the power consumption of the FRUs in a shelf, such that the overall parameters do not depart from the allowed operational envelope.

\begin{figure}[htbp]
    \centering
    \includegraphics[width=0.95\linewidth]{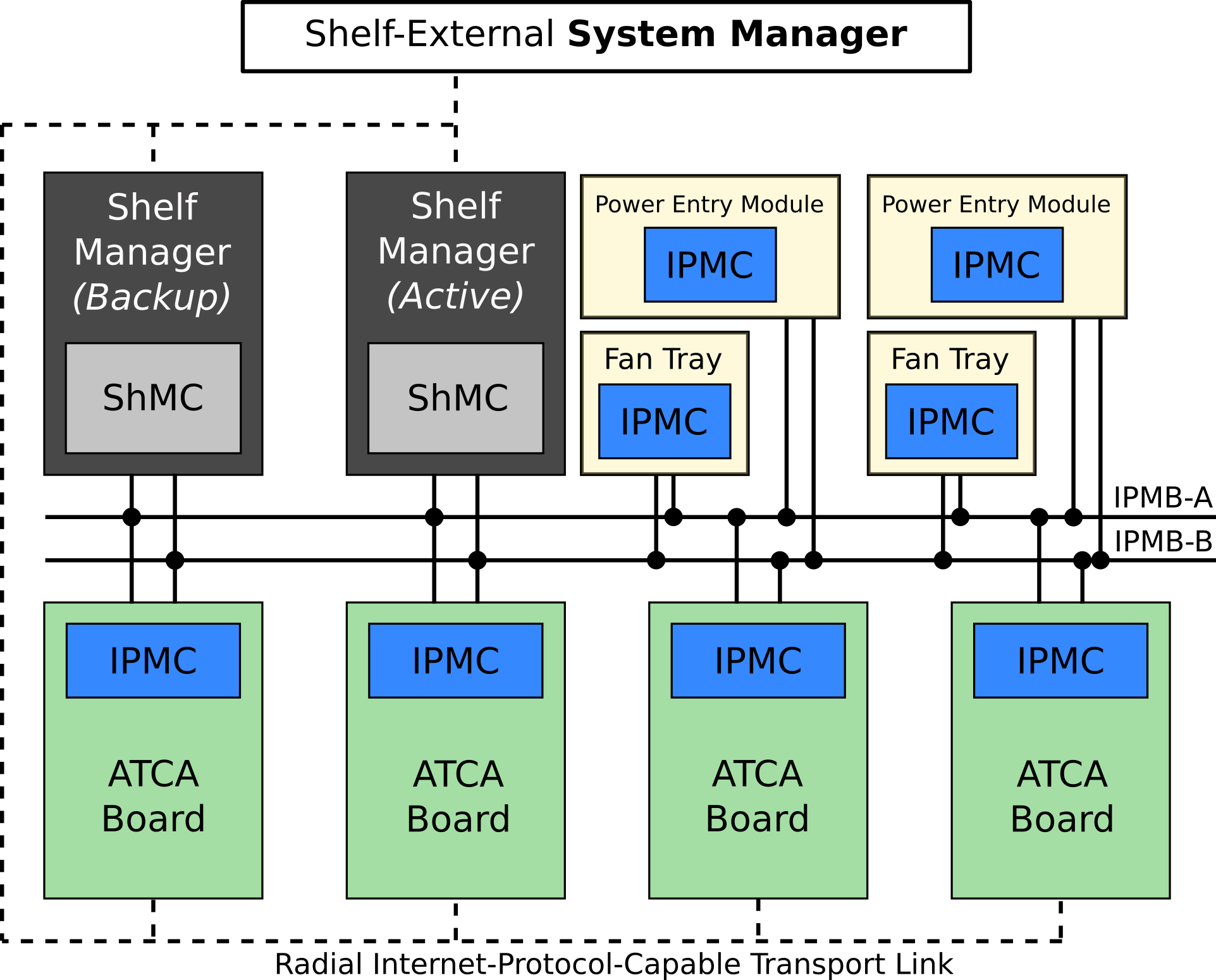}
    \caption{\label{fig:atca_management_arch} ATCA HPM Architecture. Adapted from \cite{picmg_3_0}.}
\end{figure}

As shown in Fig. \ref{fig:atca_management_arch}, the HPM system is composed of IPMCs that manage one or more FRUs, one or two Shelf Management Controllers (ShMC) for each shelf, and an optional external System Manager. The System Manager is a global high-level controller that manages ShMCs in multiple shelves connected by a network; the ShMC is a device that orchestrates the behavior of all the IPMCs running inside the FRUs hosted on its shelf; and the IPMC is a controller local to each FRU that is responsible for controlling all aspects specific to the FRU operational state and providing real-time hardware status and sensor information to the ShMC. The communication between FRUs and ShMCs takes the form of messages based on the Intelligent Platform Management Interface (IPMI) \cite{ipmi_spec} \cite{ipmb_spec} protocol. In the specific application to ATCA, the IPMI protocol is extended through the addition of remote board control, fault detection and fault management functions. 

\subsection{IPMC: Hot-swap and other functions}
One of the main roles of the IPMC is the management of the hot-swap operation, where the board is activated or deactivated in a graceful way, assisted by the ShMC. The procedure starts with the insertion of a board into a shelf, the IPMC is powered immediately, informing the ShMC about its presence. The board activation request from the user is signalled by locking the mechanical handle on the front plate of the board, which triggers a switch. The IPMC then sends to the ShMC information about its sensors (like name, units, conversion constants, thresholds and many others), board identification and power requirements. The ShMC then evaluates the power budget in the shelf and may or may not authorize to power ON the board. If power-ON permission is granted, the IPMC follows the specific steps needed to bring the electronics on the board to the active state, for example booting an operating system on a processor. In a similar fashion, when the front handle is unlocked the IPMC begins the board-specific procedure to shut down gracefully the electronics in the board, coordinating with the ShMC in doing so.

Considering the hot-swap as an example, it is clear that the main reason for the IPMC to exist in the standard, is that it exposes a standard abstract interface for the management of the board, hiding the board-specific details from the ShMC. Thanks to this standard interface, the ShMC can be designed generically and without the need for prior knowledge of the details of boards installed into the shelf. This also means that the IPMC needs to be specifically customized for the board it is designed to run in, either by configuring its generic firmware through scripts, or by designing a firmware tailored for this purpose. Other functions of the IPMC include declaring the list of sensors available on the board to the ShMC, reading them out and transmitting their readings to the ShMC, which is the device tasked with the ultimate decision on whether to command the shut down a FRU in the shelf.

\begin{figure}[htbp]
    \centering
    \includegraphics[width=0.80\linewidth]{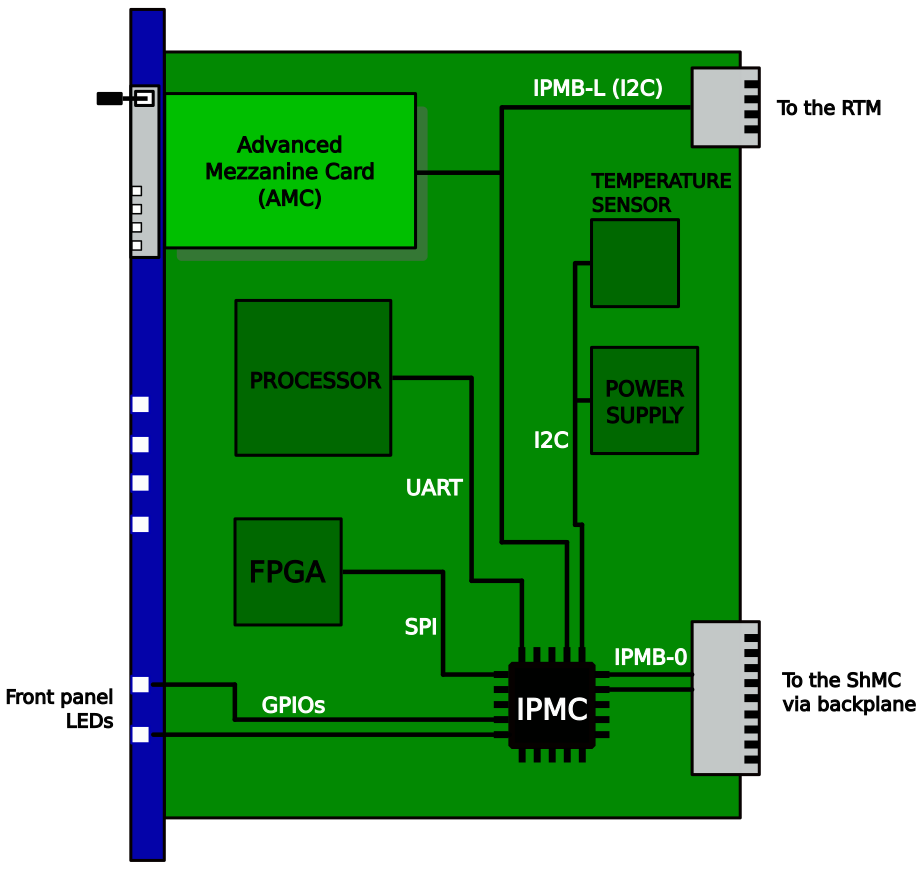}
    \caption{\label{fig:atca_card_schematic} Example scheme of an ATCA electronic card hosting expansion boards.}
\end{figure}

ATCA boards can host expansion boards (see Fig. \ref{fig:atca_card_schematic}), which can be Advanced Mezzanine Cards (AMCs)\cite{amc_spec} - meant to be inserted into front-facing slots of the main board - and Rear Transition Modules (RTMs)\cite{picmg_3_0, irtm_spec, rtm_conn_spec} - hosted in an optional slot on the rear of the backplane. When expansion boards are used, the IPMC operates as the managing controller for their operation, in a similar way as the ShMC manages the FRUs in a shelf. The communication between IPMC and the expansion boards takes place via a local IPMI bus (IPMB-L) - with electrical characteristics similar to a single IPMB-0 channel - and a number of status and control signals.

\section{OpenIPMC}
OpenIPMC is a piece of portable software implementing the behavior of an IPMC. It stems from a collaboration between São Paulo Research and Analysis Center (SPRACE) and the Karlsruhe Institute of Technology (KIT) on development of electronics for the Phase-2 upgrade of the CMS experiment. The development of OpenIPMC is the evolution of a previous project led by SPRACE researchers in collaboration with Fermilab, in which the collaborators successfully developed the IPMC \cite{ramalho_ipmc_pulsa2b}\cite{paiva_thesis} for the Pulsar2b ATCA board \cite{Ajuha:2017frj}. The free and open-source nature of OpenIPMC helps in the customization and debugging of the firmware running on prototype ATCA boards during their development and commissioning, and allows adapting it to the operational needs as they evolve over the expected long period of operation. 

The software is written in C language, based upon the FreeRTOS free and open-source real-time operating system \cite{FreeRTOS} and targeting embedded Microcontroller Units (MCUs) and systems-on-a-chip (SoCs). Thanks to the very wide support of FreeRTOS across different hardware manufacturers, OpenIPMC can easily be ported to any MCU/SoC supported by FreeRTOS, provided that the device is equipped with sufficient amount of resources to run the code and with enough I/O peripherals to interface the microcontroller to the ATCA backplane and local board functions. In our tests we estimated the size taken by OpenIPMC by observing the size increment experienced by a basic firmware when OpenIPMC was included (table \ref{tab:openipmc_size}). From the table we can also observe that, while OpenIPMC requires just a few tens of kiB, there are large variations in the overall firmware size across different microcontrollers and SoCs, likely due to differences in the board support package implementations. We have not yet optimized the stack size for the OpenIPMC real-time tasks, which will be described in the next paragraphs. The sizes presented in the table include instructions, core data and the RAM reserved for the heap working space.

\begin{table}[h!]
\caption{\label{tab:openipmc_size} Estimated size of OpenIPMC in firmware and total size of firmware for different architectures.}
\begin{center}
\begin{tabular}{ c|c|c }
 \normalsize Device & \normalsize OpenIPMC size (kiB) & \normalsize Total size (kiB) \\  
 \hline
 \normalsize ZynqUS+ & \normalsize 55 & \normalsize 218 \\ 
 \normalsize ESP32   & \normalsize 33 & \normalsize 93 \\ 
 \normalsize STM32   & \normalsize 40 & \normalsize 75
\end{tabular}
\end{center}
\end{table}

\begin{figure}[htbp]
    \centering
    \includegraphics[width=0.95\linewidth]{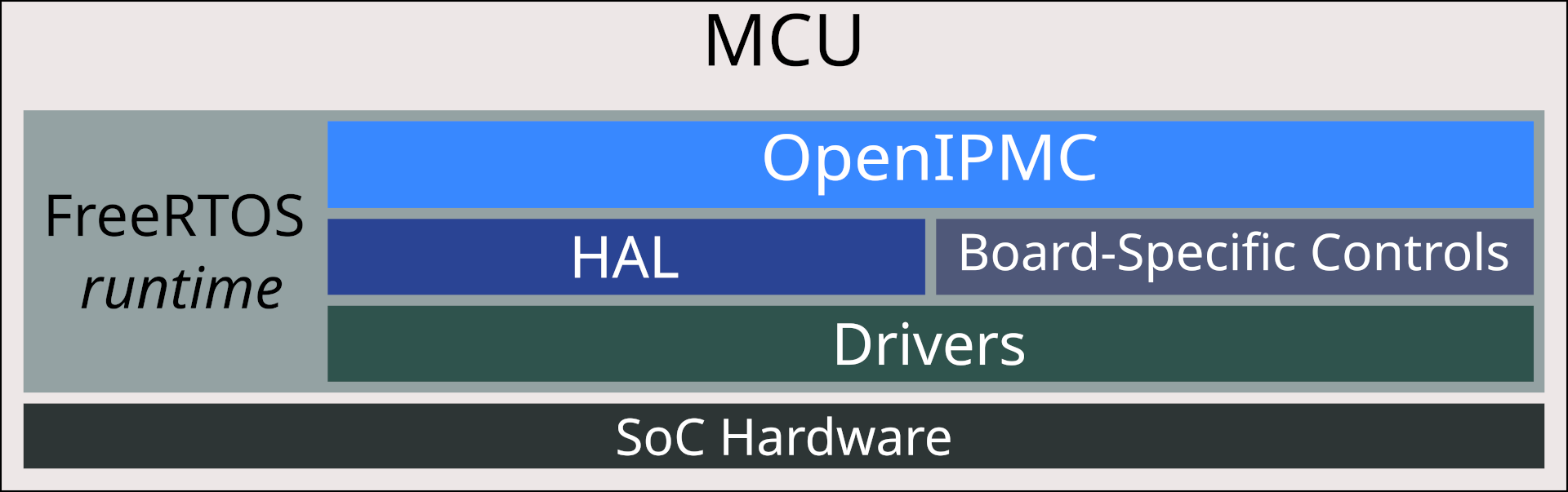}
    \caption{\label{fig:open_ipmc_in_mcu_system} Schematic of the relationships between the core of OpenIPMC, its HAL and the hardware-specific drivers in the context of the FreeRTOS runtime, and the hardware peripherals of the SoC/microcontroller.}
\end{figure}

In the case of MCUs which do not ship with enough hardware peripherals to cover all the needed $\text{I}^{2}\text{C}$ channels, communication can be established by  pairs of software-driven GPIOs to emulate such peripherals. Still, this operation can be rather CPU-intensive and therefore should be avoided on very busy channels when possible. When choosing a microcontroller or SoC for IPMC applications, we recommend the use of devices with at least three $\text{I}^{2}\text{C}$ dedicated peripherals (either hard in-silica cores or soft cores in the programmable logic of an FPGA), with two to be to used in IPMB-0 communication and at least one to control local devices such as temperature and current sensors. In cases where the board can host AMCs or RTMs, an additional hardware peripheral should be dedicated to the operation of the IPMB-L interface. OpenIPMC achieves portability thanks to a stringent separation between its core behavioral code and the accessory interface to the underlying microcontroller drivers and hardware. This is accomplished through a \emph{Hardware Abstraction Layer} (HAL) and \emph{board-specific control} functions (as shown in Fig. \ref{fig:open_ipmc_in_mcu_system}).  

\subsection{Running OpenIPMC in FreeRTOS}
OpenIPMC was designed to be integrated in a wider microcontroller firmware according to the needs of the developer and, therefore, it was a natural choice to adopt a multi-task real time operating system such as FreeRTOS as underlying infrastructure. The core of OpenIPMC runs as a collection of FreeRTOS tasks (described in the following paragraphs) running in parallel and interacting via thread-safe queues and semaphores, with no requirement for exclusive access to the processor. Hence, the developer has the freedom to extend the functions of the firmware by adding independent tasks excluded from the OpenIPMC execution flow. The FreeRTOS scheduler\cite{FreeRTOS} allows to set the priority of each task, such that the critical ones can be guaranteed to execute with low latency, unimpeded by low-priority tasks. OpenIPMC executes in parallel the tasks listed below.

\begin{figure*}[t]
    \centering
    \includegraphics[width=0.8\textwidth]{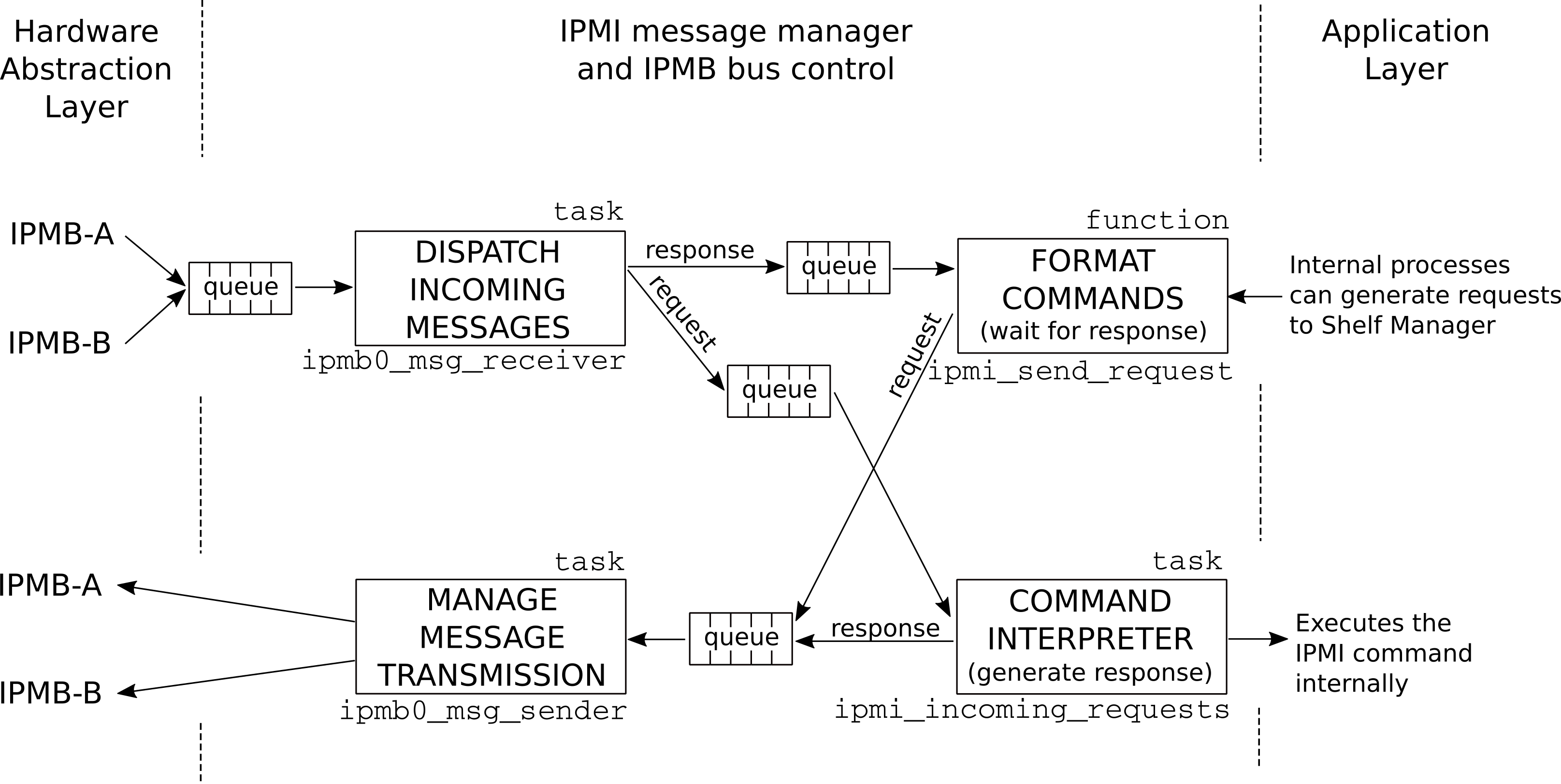}
    \caption{\label{fig:ipmc_msgs_fluxogram} IPMI message transactions in OpenIPMC. ``Hardware Abstraction Layer" represents the collection of adapter functions between OpenIPMC and the I$^2$C driver available for the desired platform. ``Application" Layer represents the set of functions responsible to execute the IPMI commands coming from ShMC. Processes in this layer also can generate requests to the ShMC by calling the proper API.}\label{fig:ipmc_msgs_fluxogram}
\end{figure*}

\textbf{\texttt{ipmb0\_msg\_receiver\_task}}
Listens for incoming messages on IPMB-A and IPMB-B. It performs basic verification on the message, checking its checksum and the type of the message (request or response). The task forwards the message to the proper queue according to its type.

\textbf{\texttt{ipmb0\_msg\_sender\_task}}
Processes messages outbound to the IPMB-A and IPMB-B channels. The task pops messages from a specific output queue, chooses the IPMB channel to use, sends the message and manages retries in case of transmission failures. 

\textbf{\texttt{ipmi\_incoming\_requests\_task}}
Is responsible for generating responses for received requests, triggering the proper processes to do so. This task interacts with both previously described \texttt{ipmb\_0\_msg} tasks.

\textbf{\texttt{fru\_state\_machine\_task}}
Drives the FRU state transitions, such as responding to the \textit{hot-swap} events and triggering the board-specific activation/deactivation routines.

\textbf{\texttt{ipmc\_handle\_switch\_task}}
Periodically samples the state of the front-plate handle and triggers \textit{hot-swap} events. This task interacts with \texttt{fru\_state\_machine\_task}.

\textbf{\texttt{ipmc\_blue\_led\_blink\_task}}
Controls the blinking time of the blue-LED, a device mandated by the PICMG standard in the front panel to indicate the power status of the board. This task polls the current state in \texttt{fru\_state\_machine\_task}.

The interplay between tasks responsible for the IPMI message transactions described above and the data flow between them are schematized in Fig. \ref{fig:ipmc_msgs_fluxogram}.

We chose to use FreeRTOS as a base for our software because it is a mature and widely-used real-time operating system designed to be robust, with a tiny footprint, and a wide range of supported devices\cite{FreeRTOS}. Among the safety features offered by this Real-Time Operating System (RTOS), we employ the stack overflow detection feature to evaluate the stability of our application, and we have the possibility to trigger a global reset of the MCU upon such an event. Furthermore, OpenIPMC has been designed to avoid the use of dynamic memory allocation after its initialization. As a good practice, all tasks and other FreeRTOS objects (queues, semaphores, etc) are allocated just once, at startup, and heap over-run events are set to be logged if they occur. The few globally-accessible symbols are protected through the use of mutexes against racing conditions.

\subsection{OpenIPMC Hardware Abstraction Layer}

\begin{minipage}{0.90\columnwidth}
\begin{lstlisting}[language=C, caption={A function implementing the blue-LED state change on a specific hardware. This \texttt{ipmc\_ios\_blue\_led} prototype is available as an example in the OpenIPMC code.}, label={lst:blue_led_callback}]
// The pointer to this function will be
// registered into the OpenIPMC HAL as the
// implementation of the blue-LED state change.
void ipmc_ios_blue_led_set(int blue_led_state)
{
  if( blue_led_state == 1 )
    gpio_driver( BLUE_LED_PIN_NUMBER, SET_TO_HIGH );
  else
    gpio_driver( BLUE_LED_PIN_NUMBER, SET_TO_LOW );
}
\end{lstlisting}
\end{minipage}
\vspace{3mm}

The HAL provides OpenIPMC with an interface to the hardware (IMPB-A, IMPB-B, hardware address pins, status LEDs, handle switch, etc.) that is prescribed in the PICMG standard to be present in every ATCA board and is fundamental for OpenIPMC operation. This HAL is composed of a set of functions that call the drivers used to access the relevant peripherals in the microcontroller. These functions take the form of declared - but undefined - functions in the base software release of OpenIPMC, and must be implemented by the developer of a new board to fit the specific hardware interface. As an example, turning ON the blue-LED, could be accomplished through a GPIO being flipped to the HIGH or LOW status, or be controlled by an on-board device like an I/O Expander attached to a local I$^2$C bus. The idea behind the OpenIPMC HAL is to give the board developer the freedom to choose how to turn ON the blue-LED in hardware and implement the corresponding function in software to do so. In Listing \ref{lst:blue_led_callback} we show as an example the code needed to operate a GPIO-driven blue status LED in the front panel.

\subsection{Board-Specific Controls}
The interaction between OpenIPMC and the electronics in the payload (that is, the electronics responsible for the main functions of the ATCA board) takes place through an API, which accepts the registering of callbacks to functions managing the various operations to be performed on the payload. The choice of using an implementation based on function callbacks is justified by the fact that boards are designed with great variety in terms of functions and components, additionally the PICMG standard makes no prescription on this aspect of the board design. Through the API the user can implement payload control routines that fit the hardware design of his choice. We chose to call this layer, made of callable hardware interface functions, the "board-specific controls" (see Fig. \ref{fig:open_ipmc_in_mcu_system}).

Among the classes of operations that need to be implemented in the IPMC according to the PICMG standard are the ones relative to Power Management and Sensor Readings. Power Management refers to the activation, deactivation, reset and the regulation of the power draw of the different circuits present on the payload. In OpenIPMC this management is performed by the state machine implemented in \texttt{fru\_state\_machine\_task}, which calls a number of user-defined board-specific controls. Different states (M0, M1, ...) of the state machine represent the different power states of the payload during activation and deactivation operations. Since the ShMC centralizes the management of power allocation for all the ATCA boards in the shelf, the activation/deactivation process involves negotiations between OpenIPMC and the ShMC through the IPMB bus. The other operations on the payload specified by the PICMG standard, like \textit{Cold Reset} and \textit{Warm Reset}, can be also implemented through board-specific controls.

The IPMC must be able to collect sensor readings from the payload and send them to the ShMC, formatted in accordance with the IPMI specifications. Similarly to the case of Power Management described above, it is the responsibility of the user to provide a function callback such that OpenIPMC can read a sensor value using the correct procedure and protocol (for example, by accessing an SPI or $\text{I}^{2}\text{C}$ register, writing and reading a GPIO, using a lookup table to interpret data, etc.). A sensor reading is generally triggered by a request sent by the ShMC. OpenIPMC receives this request on the IPMB bus, interprets it and executes the callback associated with the reading of that specific sensor, and finally sends the value and sensor status to the ShMC.

According to the standard, the IPMC is also responsible for sending to the ShMC information about each sensor such as sensor type, measurement units, linearization parameters, threshold values, accuracy, a string containing its name, and other information, which is collected into a data structure called the \emph{Sensor Data Record} (SDR)\cite{ipmi_spec}. This is generated by the IPMC at startup and transmitted to ShMC during the activation process. OpenIPMC provides an API to create SDRs for the board sensors and automatically manages their transmission to the ShMC.

The implementation of the Board-Specific controls and of the HAL layers strongly depend on the specific hardware being targeted and on its driver interface, as designed by the hardware manufacturer. Due to the wide variety of microcontrollers supporting FreeRTOS from different manufacturers, the implementation and debugging of these layers should be tailored to the target hardware on a device-by-device basis. We believe that the design of this layer should be left in control of the user to match his specific needs.

\subsection{Licensing and code distribution}
OpenIPMC is released under Mozilla Public License 2.0\cite{mpl2.0}, a license which allows  of the open-source code to be statically linked, as it is common when building microcontroller firmware binaries. The code is currently publicly available on Gitlab.com\cite{openipmc_repo}.

\section{Development and Testing Platforms}
We began the development of OpenIPMC on the AVNET Ultra96 board \cite{ultra96}, establishing early the separation between behavioral code, HAL and the board-specific controls described before. This separation was of great help to port the code to a number of different architectures. With the exception of the Trenz-Serenity one, all tests were performed at São Paulo State University using an ATCA horizontal shelf manufactured by Comtel (model CO6B-6U-FM40X) with 6 slots and two redundant ShMM-700R Shelf Management Controllers by PigeonPoint (running firmware version 3.6.1.3).

\begin{figure}[htbp]
    \centering
    \includegraphics[width=0.95\linewidth]{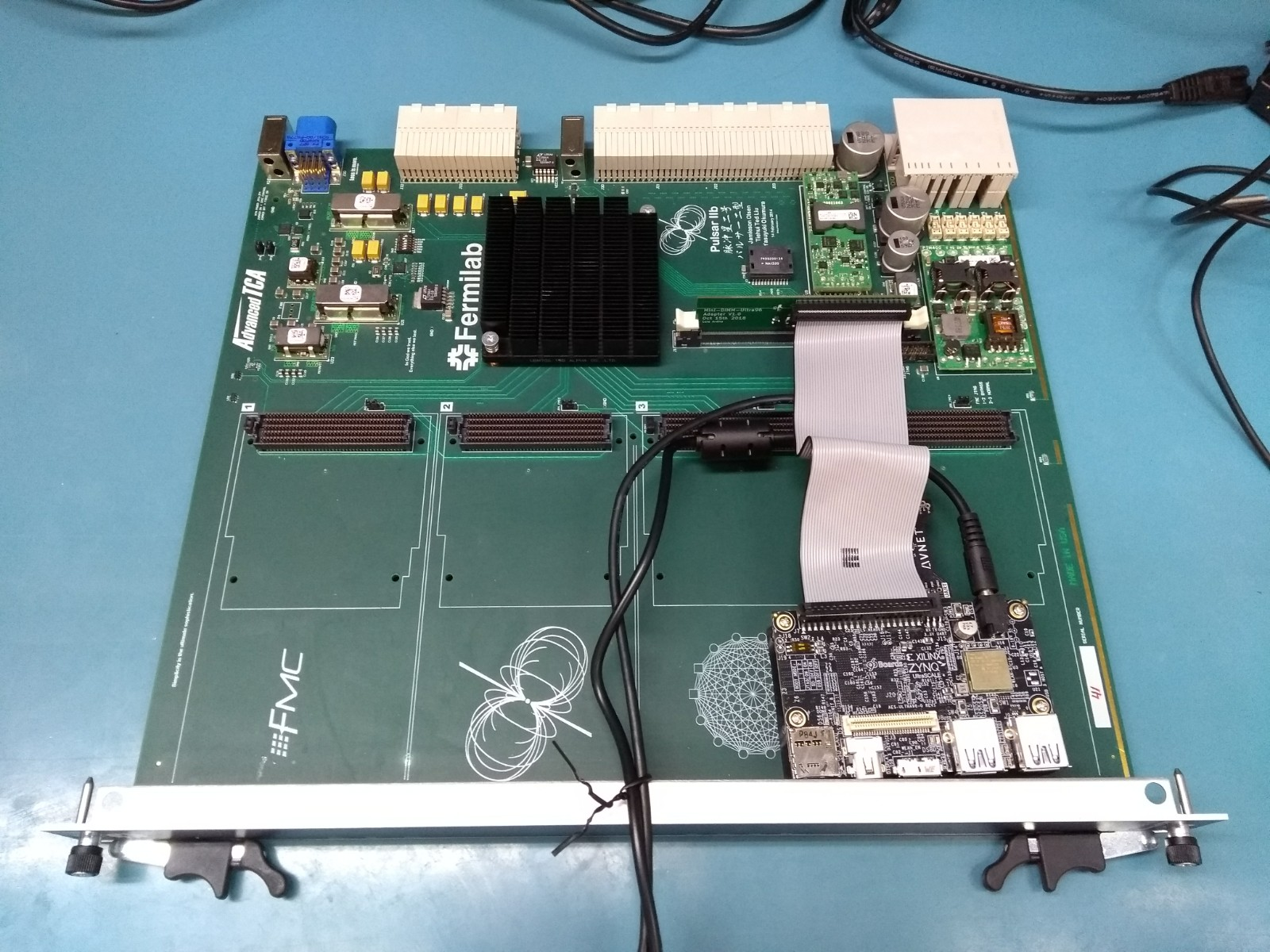}
    \caption{\label{fig:pulsar_ultra96_top} Development setup with the Ultra96 sitting on top of the Pulsar-2b.}
\end{figure}

\subsection{Ultra96-Pulsar2b Development Platform}

In this setup the Ultra96 acts as an IPMC for a Pulsar-2b board. The ZU3EG SoC on the Ultra96 board belongs to the ZynqUS+ EG family\cite{xilinx_zynq_ds,xilinx_zynq_trm}, containing four ARM A53 high-performance application processors (APU) and two ARM R5 real-time processors (RPU). APU and RPU are independent processing units within the same SoC package which can run different operating systems. The FreeRTOS instance hosting the OpenIPMC tasks runs on the RPU, leaving the APU free to run a Linux-based operating system. The communication between the Ultra96 and the ATCA backplane takes place through an adapter board, which fits into the Mini-DIMM slot normally used by the Pulsar-2b to host its IPMC board. This adapter board exposes on a 2\.mm  header the IPMB-A and IPMB-B buses, the hardware address lines, the blue-LED control and the handle switch state line, such that they can be connected to the pins of the Ultra96 through a flat cable (Fig. \ref{fig:pulsar_ultra96_top}).

In tests performed on this setup, OpenIPMC correctly executes its management tasks, such as the activation and deactivation triggered by the handle switch, and the declaration and read-out of a dummy sensor to the ShMC. In this setup, both power and sensor management are simulated, since the current version of the DIMM adapter does not provide access to the sensors $\text{I}^{2}\text{C}$ bus and the power supply controls of the Pulsar-2b. Furthermore, using this platform, OpenIPMC was used to trigger the boot sequence of a CentOS Linux distribution \cite{centos-zynq-cern} on the APU of the ZynqUS+.

\subsection{Trenz-Serenity Development Platform}

\begin{figure}[htbp]
    \centering
    \includegraphics[width=0.75\linewidth]{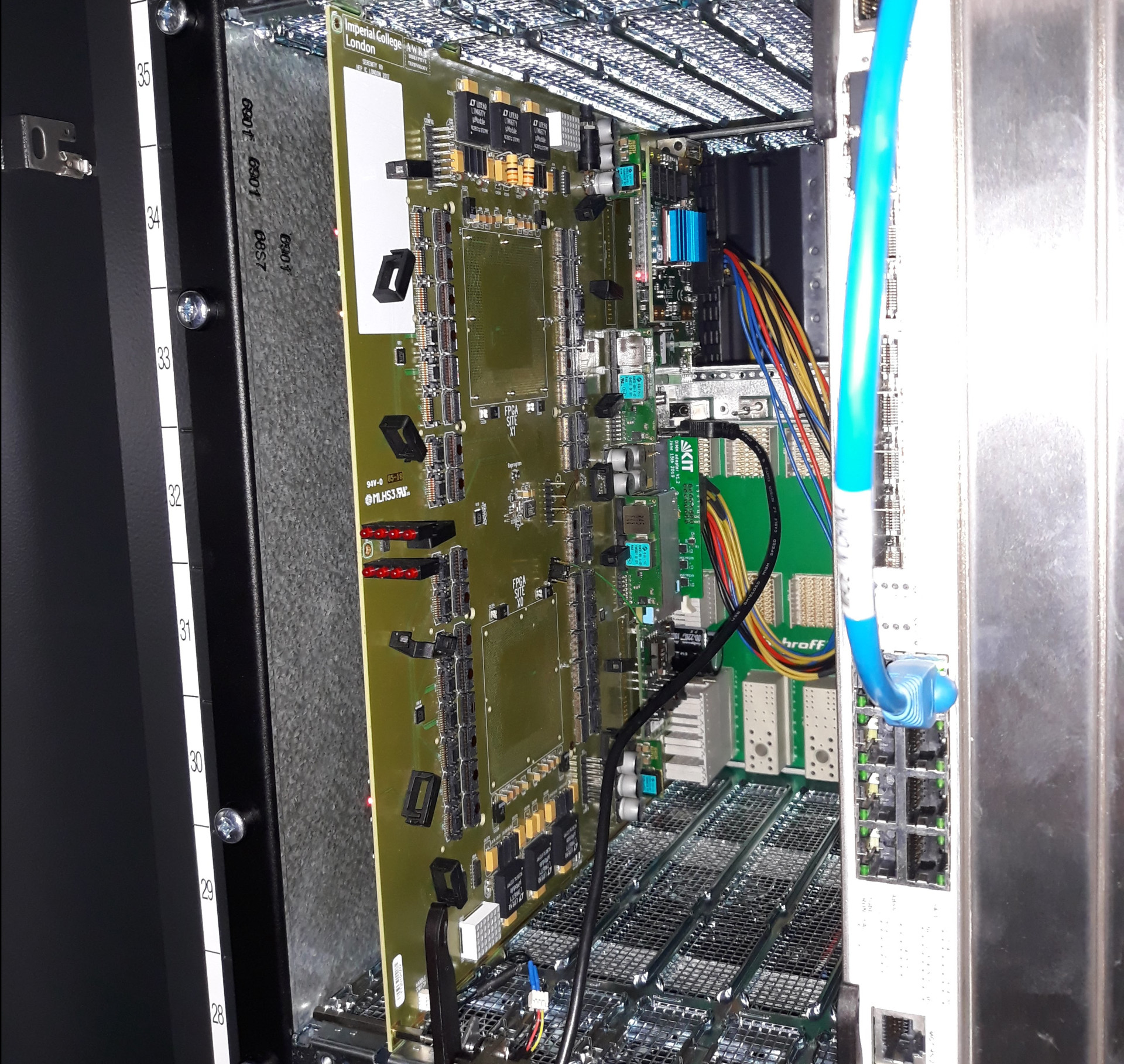}
    \caption{\label{fig:zynqmp_serenity} Trenz-Serenity setup in a vertical shelf at KIT.}
\end{figure}

The Trenz-Serenity Platform is the platform currently being used at KIT for the development of a centralized management architecture based on ZynqUS+\cite{atca-zynqmp-ipmc}. Its main components are the Serenity ATCA Carrier Card \cite{Rose:2019oiy} and a Trenz TE0803 module \cite{trenz_te0803} hosted on a custom adapter card (\textit{Trenz Adapter}), which allows the TE0803 to fit into the slot - originally designed to accommodate a COMExpress industrial computer \cite{com_express_picmg} - present on the Serenity board (Fig. \ref{fig:zynqmp_serenity}). Additionally, the Trenz Adapter interconnects the TE0803 module to the IPMC slot on the Serenity board through a ribbon cable and an adapter Mini-DIMM board, allowing the ZynqUS+ on the TE0803 module to perform the role of an IPMC.

Similarly to the case of the Ultra96-Pulsar2b platform, the Trenz TE0803 module hosts a ZynqUS+ EG device. OpenIPMC runs on the ARM Cortex-R5 cores and the IPMB channels are implemented using both I$^2$C hard peripherals available on the ZynqUS+ \textit{Processing System} (PS). However, this test setup presents a number of significant differences compared to the Ultra96 case and the OpenIPMC HAL subsystem has been modified accordingly. For example, due to the different signal routing on the Trenz module some critical signals (Hardware Address, blue-LED, Handle\_Switch and 12V\_Enable) have been routed to PCA9557 IO expanders controlled by an I$^2$C master in the ZynqUS+. Since both I$^2$C channels available on the ZynqUS+ PS are already allocated to the IPMB buses, to drive the expander we use a software-emulated I$^2$C master, where two GPIO signals are controlled by a driver to behave as the SDA and SCL lines of an I$^2$C peripheral.

On this setup, OpenIPMC has shown to correctly perform its management tasks, including turning ON the main board power supply, managing sensors and triggering the boot of a CentOS Linux distribution running on the APU of the ZynqUS+.

\subsection{ESP32-Pulsar2b Development platform}
To demonstrate the hardware independence of the core behavioral code of OpenIPMC we performed an exercise, porting OpenIPMC to an architecture significantly different to the ZynqUS+ SoC used in the two development setups described earlier, and measuring the human effort needed to complete the porting. We chose as platform the Espressif ESP32 microcontroller, which is based on a Harvard-architecture Tensilica Xtensa LX6 processor, significantly different from the ARM R5 cores on the ZynqUS+. The ESP32 microcontroller is an affordable yet powerful device designed for wireless IoT applications \cite{esp32_datasheet} \cite{esp32_tech_ref_man}.

\begin{figure}[htbp]
    \centering
    \includegraphics[width=0.90\linewidth]{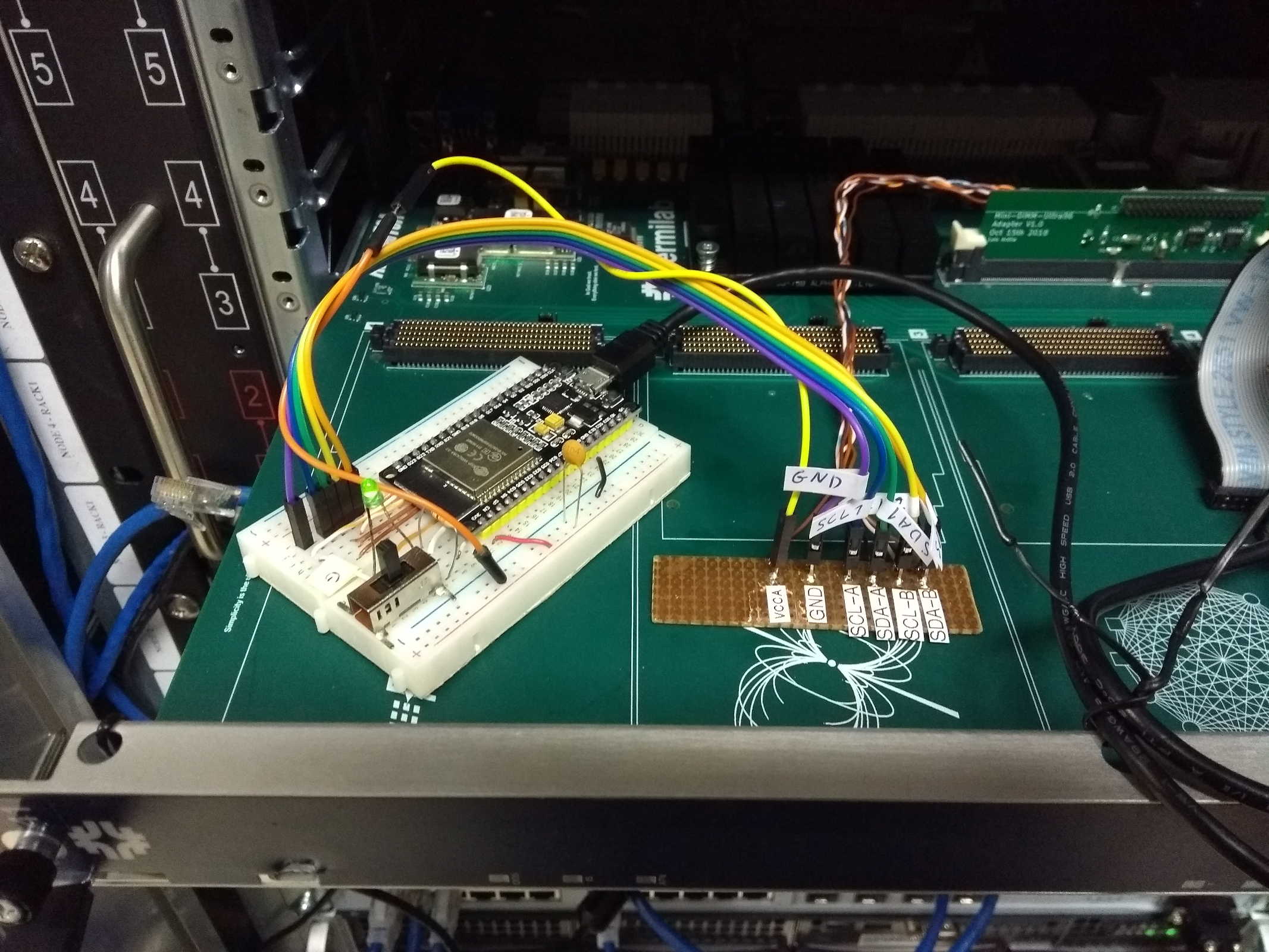}
    \caption{\label{fig:ipmc_esp32_2} ESP32-Pulsar2b development platform at SPRACE in São Paulo.}
\end{figure}

We assembled the platform for this demonstration starting from the Ultra96-Pulsar-2b setup and swapping the Ultra96 with a breadboard on which we installed an ESP32 development board, as shown in Fig. \ref{fig:ipmc_esp32_2}. In this setup, the only signals connected are the IPMB buses (allowing communication to the ShMC), the handle switch (which was emulated by a simple bi-stable switch), and the blue-LED (a simple through-hole LED), the last two being installed on a breadboard. For simplicity the Hardware Address, being a set of static signals bound to simple GPIO pins, was not read out and its expected values were trivially hard-coded into the HAL.

Porting OpenIPMC to this platform required just 3 person-weeks, which we identify as a success. Most of the time was spent in circumventing an inflexible implementation of the I$^2$C multi-master mode in the ESP32 IDF board support package, while porting of the core OpenIPMC code required very little effort. The difficulty arises from the I$^2$C driver when the device is in slave mode listening for messages: in the ESP32 the driver expects the developer to know in advance the size of each future incoming message, which is not the case for IPMB communication, where messages from the master can have variable length. We tested the behavior of the IPMC and its interaction with the ShMC and found it to work as expected.

\begin{table*}[t]
\caption{Comparison of peripheral usage between the different test setups. PS = Processing System and PL = Programmable Logic in ZynqUS+.}
\label{tab:setups_comparison}
\centering
\begin{tabular}{l|c|c|c|c}

                                                                                            & Ultra96-Pulsar2b & Trenz-Serenity                                                                          & ESP32-Pulsar2b       & STM32-Pulsar2b \\ \hline
\multicolumn{1}{l|}{SoC/microcontroller}                                                    & ZynqUS+ ZU3EG    & ZynqUS+ ZU4EG                                                                           & ESP32                & STM32H745      \\ \hline
\multicolumn{1}{l|}{Architecture}                                                           & ARM Cortex-R5    & ARM Cortex-R5                                                                           & Tensilica Xtensa Lx6 & ARM Cortex-M7  \\ \hline
\multicolumn{1}{l|}{IPMB-A}                                                                 & I$^2$C-PS        & I$^2$C-PS                                                                               & I$^2$C               & I$^2$C         \\ \hline
\multicolumn{1}{l|}{IPMB-B}                                                                 & I$^2$C-PS        & I$^2$C-PS                                                                               & I$^2$C               & I$^2$C         \\ \hline
\multicolumn{1}{l|}{\begin{tabular}[c]{@{}l@{}}Hardware \\ Address (bit 0)\end{tabular}}    & GPIO-PL          & {\begin{tabular}[c]{@{}c@{}}PCA9557 using software-emulated \\ I$^2$C on GPIO-PS\end{tabular}} & GPIO                 & GPIO           \\ \hline
\multicolumn{1}{l|}{\begin{tabular}[c]{@{}l@{}}Hardware \\ Address (bits 1-7)\end{tabular}} & GPIO-PS          & {\begin{tabular}[c]{@{}c@{}}PCA9557 using software-emulated \\ I$^2$C on GPIO-PS\end{tabular}} & GPIO                 & GPIO           \\ \hline
\multicolumn{1}{l|}{Blue\_LED}                                                               & GPIO-PL          & {\begin{tabular}[c]{@{}c@{}}PCA9557 using software-emulated \\ I$^2$C on GPIO-PS\end{tabular}} & GPIO                 & GPIO           \\ \hline
\multicolumn{1}{l|}{Handle\_Switch}                                                          & GPIO-PL          & {\begin{tabular}[c]{@{}c@{}}PCA9557 using software-emulated\\ I$^2$C on GPIO-PS\end{tabular}} & GPIO                 & GPIO           \\ \hline
\multicolumn{1}{l|}{Soft\_Reset}                                                             & -                & GPIO-PS                                                                                 & -                    & -              \\ \hline
\multicolumn{1}{l|}{12V\_Enable}                                                            & -                & {\begin{tabular}[c]{@{}c@{}}PCA9557 using software-emulated \\ I$^2$C on GPIO-PS\end{tabular}} & -                    & -              \\
\end{tabular}
\end{table*}

\subsection{STM32-Pulsar2b Development platform}

Aiming for deployment into a mass-produced, reliable and affordable microcontroller, we chose to port OpenIPMC to the STM32 family \cite{stm32_website}, which is one of the most widely used families of microcontrollers worldwide. They are available in a large variety of types optimized for a large number of application classes. Specifically, the STM32H7 family is composed of high-performance microcontrollers \cite{stm32h7xx_ref_manual}, characterized by powerful processors and a large number of IO peripherals, which makes them interesting for use in a feature-rich IPMC. On this test platform, we replaced the ESP32 board in the ESP32-Pulsar2b Development platform with the NUCLEO-H745ZI-Q development board manufactured by ST Microelectronics, which hosts a STM32H745ZIT6U microcontroller \cite{stm32h745_datasheet}. This device is characterized by a dual Cortex-M7/Cortex-M4 core, four I$^2$C, four USART and four UART peripherals, and a large number of GPIO channels.

\begin{figure}[htbp]
    \centering
    \includegraphics[width=0.90\linewidth]{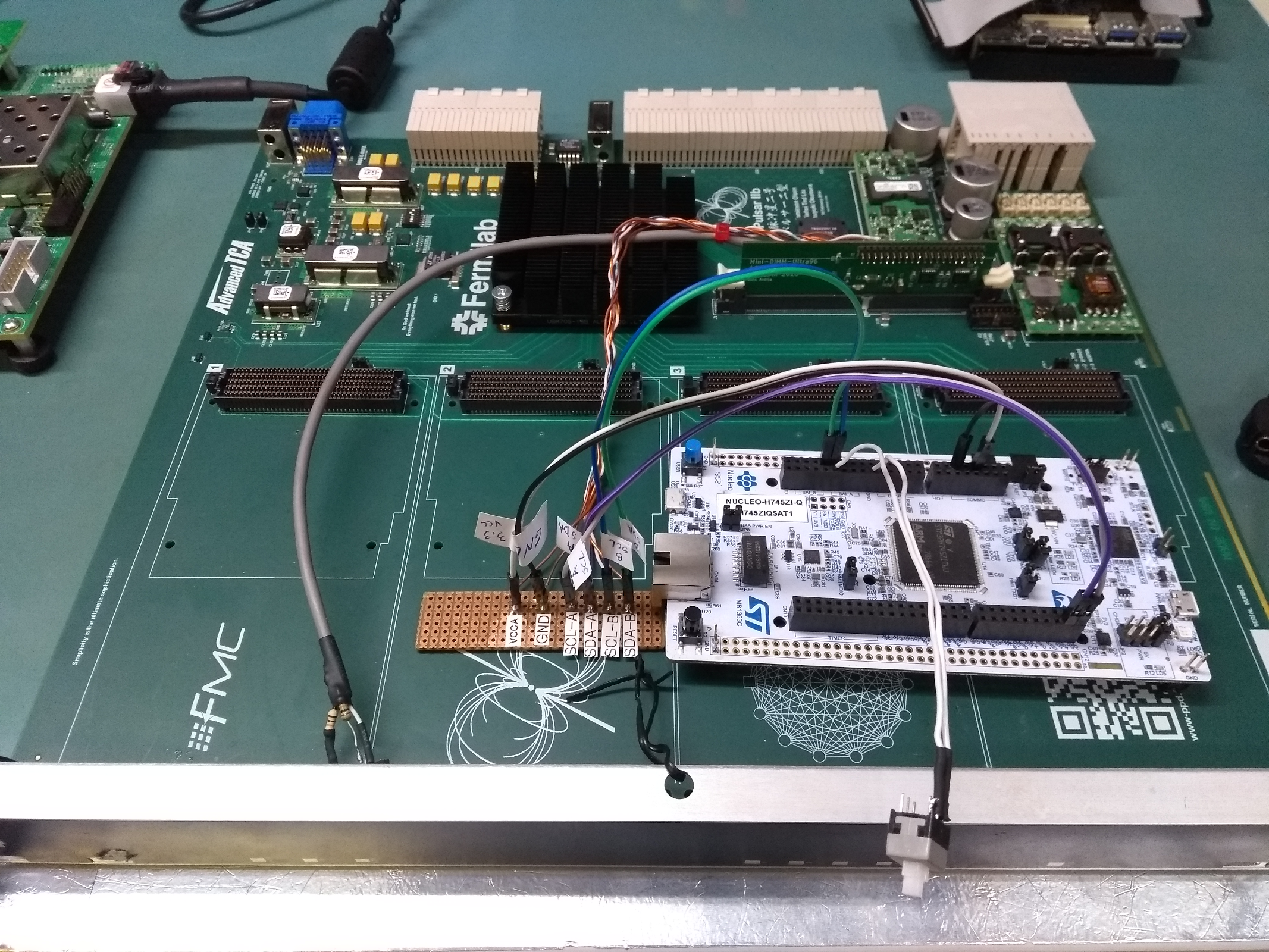}
    \caption{\label{fig:pulsar_stm32_top} STM32-Pulsar2b development platform at SPRACE in São Paulo.}
\end{figure}

\begin{figure}[htbp]
    \centering
    \includegraphics[width=0.90\linewidth]{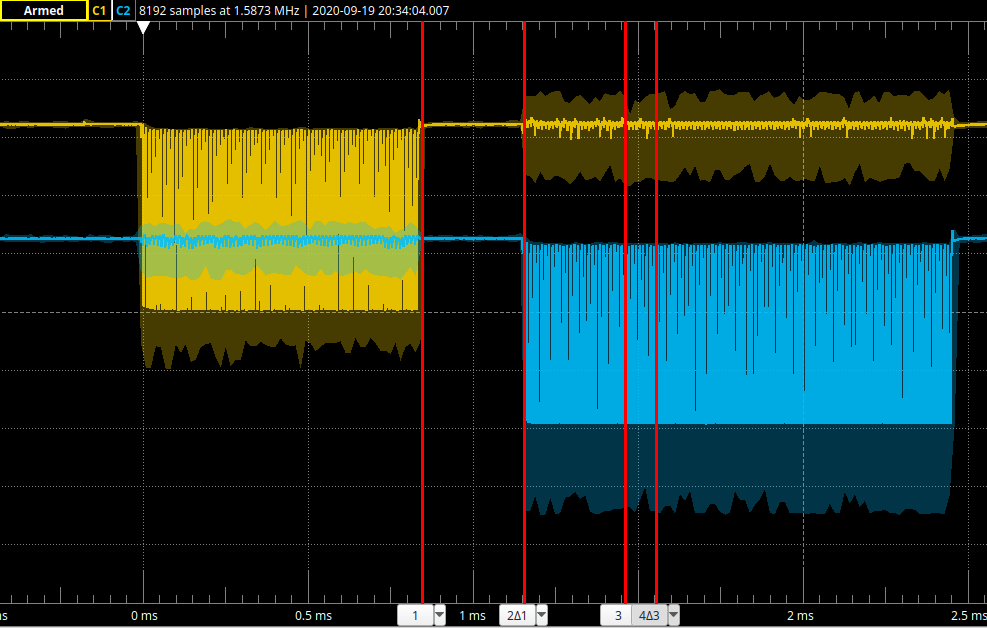}
    \caption{\label{fig:ipmb_communication_esp32} Example of IPMB communication: OpenIPMC is running on the NUCLEO-H745ZI-Q board, receiving a message from the ShMC on IPMB-A (activity on the left) and replying on IPMB-B (activity on the right) in around 300 $\mu$s. The apparent cross-talk is an artifact of the measurement setup.}
\end{figure}

We ported OpenIPMC on the STM32 successfully and with relatively little effort, taking around 5 person-weeks. Most of the effort was focused in circumventing an inflexible implementation of I$^2$C multi-mastering, similar to the case of ESP32. We could verify that OpenIPMC behaves as expected in its communication with the ShMC and in the activation process. In Fig. \ref{fig:ipmb_communication_esp32} an oscilloscope trace of the IPMB-A and IPMB-B SDA buses shows the exchange of messages between the IPMC and the ShMC. The transaction takes place in around 300 $\mu$s, well below the IPMI latency requirement.

\vspace{-7mm}
\subsection{Comparison between different test platforms}
In table \ref{tab:setups_comparison} we summarize the differences in the use of different peripherals to read and operate the various I/O needed by OpenIPMC. Note that \textit{Soft-Reset}  and \textit{12V\_Enable} are board-specific controls, which were only used in the Trenz-Serenity setup.

\section{Summary and outlook}
In this document we presented OpenIPMC, a software written in C for embedded microcontrollers implementing an IPMC as defined by the PICMG ATCA standard. The operation of the software has been demonstrated successfully on different hardware architectures such as ZynqUS+, ESP32 and STM32. We plan to continue our development on new hardware designs, where OpenIPMC will be given full control and monitoring duties over its hosting ATCA carrier board. We also plan to introduce support for local add-on boards, such as AMCs and RTMs. While  OpenIPMC has been conceived in a context of academic research, its full customizability make it attractive for many other applications, such as innovative industrial designs and board prototyping.

\section*{Acknowledgment}
The authors wish to acknowledge the Fundação de Amparo à Pesquisa do Estado de São Paulo for its financial support through grants number 18/18955-0 and 17/16245-3. We also want to thank the members of the CMS Phase-2 Tracker Upgrade Data Processing Systems group for the regular exchange of ideas, and in particular Dan Gastler, Gregory Iles, Eric Hazen, Peter Wittich and Mark Pesaresi for the help in defining the requirements for IPMCs used in the Phase-2 back-end boards. We would also like to thank Sthefany Fernandes de Souza for her effort in setting up and testing FreeRTOS.\\

\end{document}